\documentclass[aps,prb,superscriptaddress,twocolumn,bibliography,citeautoscript]{revtex4-2}
\usepackage{amsfonts}
\usepackage{amstext}
\usepackage{amssymb}
\usepackage{amsmath}
\usepackage[utf8]{inputenc}
\usepackage{graphicx}
\usepackage{graphics}
\usepackage{cancel}
\usepackage{color,xcolor}
\usepackage{hyperref}
\hypersetup{colorlinks=true,citecolor=blue,linkcolor=blue,filecolor=blue,urlcolor=blue,pdfstartview=FitH,pdfpagemode=UseNone}
\usepackage{lipsum}
\usepackage{soul}
\usepackage{pdfpages,pgffor}
\definecolor{boxcolor}{HTML}{108f64}
\newcommand{\orcid}[1]{\href{https://orcid.org/#1}{\includegraphics[width=8pt]{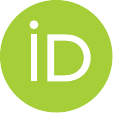}}}

\makeatletter
\AtBeginDocument{\let\LS@rot\@undefined}
\makeatother

\begin{document}
\title{Performance Assessment of Universal Machine Learning Interatomic Potentials: \\ Challenges and Directions for Materials' Surfaces} 

\author{Bruno Focassio \orcid{0000-0003-4811-7729}} 
\email{bruno.focassio@lnnano.cnpem.br}
\affiliation{Brazilian Nanotechnology National Laboratory (LNNano/CNPEM), 13083-100, Campinas, SP, Brazil}
\author{Luis Paulo Mezzina Freitas \orcid{0000-0002-1874-5612}} 
\affiliation{Brazilian Nanotechnology National Laboratory (LNNano/CNPEM), 13083-100, Campinas, SP, Brazil}
\author{Gabriel R. Schleder \orcid{0000-0003-3129-8682}}
\email[Corresponding author. ]{gabriel.schleder@lnnano.cnpem.br}
\affiliation{Brazilian Nanotechnology National Laboratory (LNNano/CNPEM), 13083-100, Campinas, SP, Brazil}
\affiliation{John A. Paulson School of Engineering and Applied Sciences, Harvard University, Cambridge, Massachusetts 02138, USA}

\begin{abstract} 
Machine learning interatomic potentials (MLIPs) are one of the main techniques in the materials science toolbox, able to bridge \textit{ab initio} accuracy with the computational efficiency of classical force fields. This allows simulations ranging from atoms, molecules, and biosystems, to solid and bulk materials, surfaces, nanomaterials, and their interfaces and complex interactions. A recent class of advanced MLIPs, which use equivariant representations and deep graph neural networks, is known as universal models. These models are proposed as foundation models suitable for any system, covering most elements from the periodic table. 
Current universal MLIPs (UIPs) have been trained with the largest consistent dataset available nowadays. However, these are composed mostly of bulk materials' DFT calculations. 
In this article, we assess the universality of all openly available UIPs, namely MACE, CHGNet, and M3GNet, in a representative task of generalization: calculation of surface energies. We find that the out-of-the-box foundation models have significant shortcomings in this task, with errors correlated to the total energy of surface simulations, having an out-of-domain distance from the training dataset. Our results show that while UIPs are an efficient starting point for fine-tuning specialized models, we envision the potential of increasing the coverage of the materials space towards universal training datasets for MLIPs. %
\end{abstract}

\maketitle

\section{Introduction}

Machine learning (ML) and artificial intelligence (AI) are rapidly changing how many aspects of our society evolve. In particular, these new approaches and techniques are added to the scientists' toolbox, and combined with traditional ones, are increasingly demonstrating novel ways to advance scientific challenges. One of the most impressive advances in recent materials science has been the use of artificial intelligence to improve and accelerate the process of materials discovery, design, and understanding \cite{Schleder2019,Butler2018, Schleder2020,Sanchez-Lengeling2018,Schleder2021}.
The field of machine learning interatomic potentials (MLIPs) has benefited the most from these advances, being currently one of the largest and most mature areas successfully using machine learning \cite{Deringer2019,Unke2021,Friederich_2021,Fedik2022}. 

MLIPs enable us to bridge the gap of computational simulations with both ab initio accuracy (typically at the DFT level of theory) and traditional force-field computational efficiency. They are capable of succeeding by going beyond traditional few-parameter analytical forms for describing the potential energy surfaces (PES) of systems. This is done by combining flexible machine-learned functional forms and algorithms with features or descriptors that are particularly adequate for these data-driven methods.
To train such flexible and higher complexity functions, high-quality and high-volume datasets are mandatory \cite{Morrow2023}, with datasets sufficiently representing the domain of interest. Even greater care is needed for generalization not only to larger simulations but also to a wide range of different systems than those of the reference dataset \cite{Li2023}, known as extensibility and transferability, respectively. \cite{Fedik2022}

Recently, a class of MLIPs employing deep graph networks \cite{Xie2018,Chen2019,Choudhary2021,Drautz2024} has improved previous results in a range of systems and benchmarks. With the use of E(3)-equivariant representations, models have achieved not only accuracy but also unprecedented data efficiency, such as the Neural Equivariant Interatomic Potentials (NequIP) \cite{Batzner2022}.
These advances led the community to envision universal MLIPs (UIPs), general models able to simulate any system across most of the periodic table \cite{Ko2023}. These universal potentials have been proposed as foundational models, a starting point from which further specialized models could be trained upon.

Chronologically, these models improved by growing both their datasets and number of parameters: M3GNet \cite{Chen2022} was trained with 188k structures and has 228k params; CHGNet \cite{Deng2023} uses atomic magnetic moments as additional representation inputs and introduced the MPtrj dataset with 1.58M structures from the Materials Project database and has 413k params; MACE MP-0 \cite{Batatia2022mace,Kovcs2023,MACE_MP-0} was also trained with MPtrj and has 4.69M params; the proprietary GNoME model \cite{Merchant2023} was trained with an active learning dataset of 89M structures starting from the Materials Project, and has 16.2M parameters; and the recent proprietary MatterSim models \cite{Lu2024_mattersim} use a dataset of 17M structures and can reach up to 182M parameters.
As usual in the deep learning community, a catalyst for advances is the creation of benchmarks to assess performance in a comparable task, the \textit{Matbench Discovery} \cite{MatbenchDiscovery} serves this purpose in the task of materials discovery for universal models.
However, the models are trained on the largest systematic DFT databases available to date, which include millions of bulk materials simulations but are currently not comprehensible in the complete materials space of all possible combinations of atoms, compositions, and structures, such as molecules, surfaces and interfaces, nanomaterials, and so on.

Here we investigate one of the main questions for users of MLIPs: given a new problem, what is the most accurate and efficient way to tackle it?
The assessment follows two steps, \textit{i)} evaluating if universal MLIPs are accurate enough for systems that are significantly different than the training dataset, with the representative example case of surfaces, and \textit{ii)} if it is better to train a specialized MLIP from scratch or to fine-tune a universal foundational model.
We show that currently available UIPs do not display sufficient zero-shot performance in this representative task. In terms of efficiency, fine-tuning UIPs can accelerate training and accuracy by incorporating alchemical transferable knowledge, thus requiring only a modest dataset to achieve sufficient accuracy in specialized tasks.

\section{Results and Discussion}

\subsection{Dataset exploration}

We start by querying the Materials Project (MP) database for data on surfaces. This data is available for unary systems \cite{Tran2016}. We gathered 1497 different surface structures that were generated from 138 different bulk systems, comprehending 73 different chemical elements.

\begin{figure}[htb!]
\includegraphics[width=\linewidth]{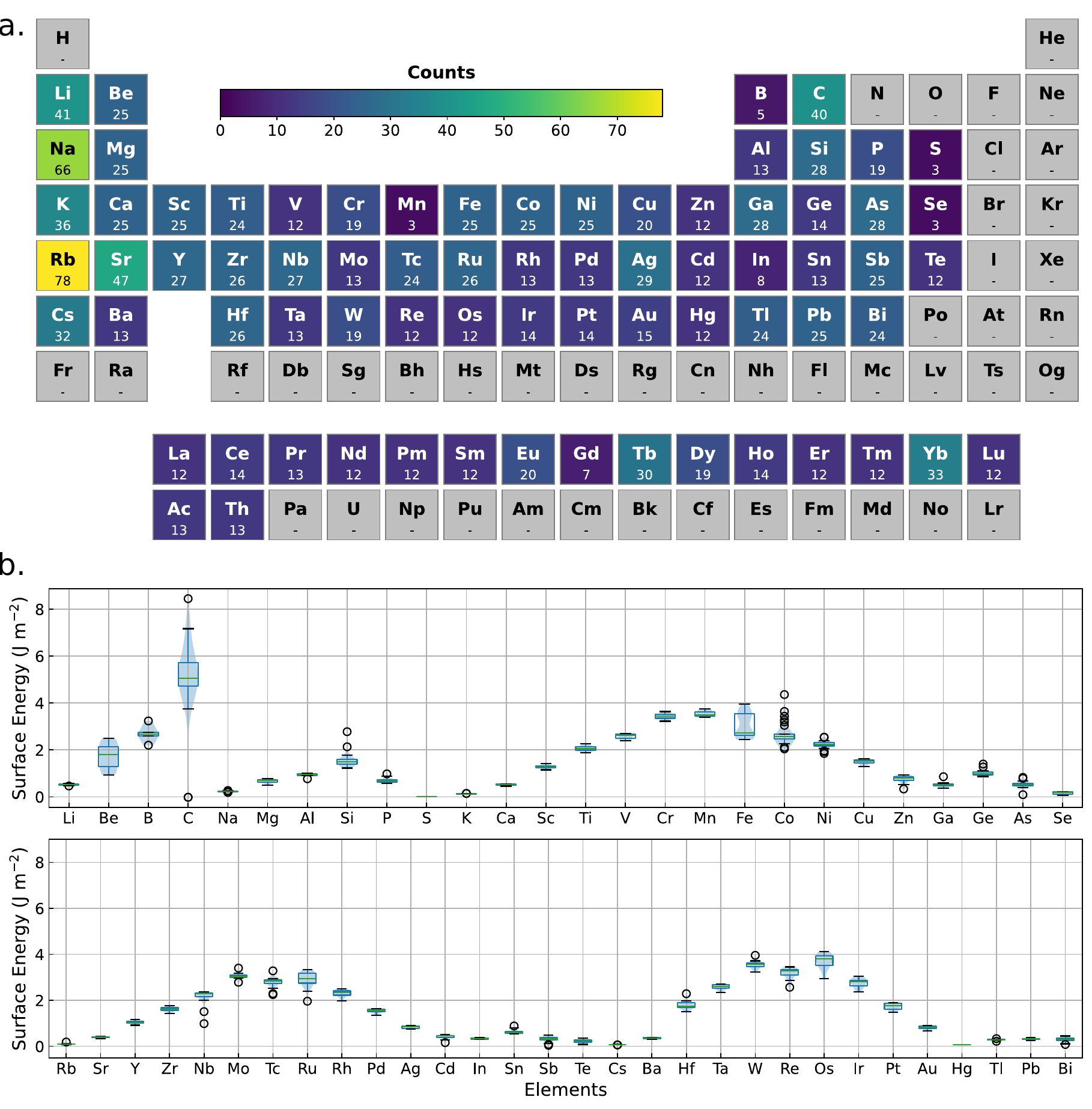}
\caption{MP surfaces dataset exploration. (a) Periodic table heatmap for the number of surfaces of each element. (b) Boxplot and violin plot for surface energy of each element within the dataset. The horizontal lines in the middle of the boxes mark the medians. The boxes are plotted from the first to the third quartile. The whiskers extend to $1.5$ times the interquartile range from the third (above) and first (below) quartile and empty circles mark the outliers (data points outside the whiskers).}
\label{fig:dataset}
\end{figure}

Figure \ref{fig:dataset} shows the distribution of surface data available on Materials Project for each element. Most materials are contemplated, however, noble gases, rare-earth elements, and also heavy synthetic elements are not present, see Fig.~\ref{fig:dataset}a. The alkali metals and alkaline earth metals are the groups with the most representation, followed by carbon materials. From Fig.~\ref{fig:dataset}b, we observe that the median surface energy increases from the edges to the center of the same period in the periodic table. The same is not observed for rare-earth elements (see the Supplemental Information). Overall the spread of surface energies is small, except for carbon materials, which also show significant outliers.

\subsection{Inference of universal machine learning atomistic potentials}

We proceed with an assessment of the universal MLIPs (UIPs) in their ability to predict the surface energy data of the materials under consideration. In this work we focus on using the following UIPs: MACE \cite{MACE_MP-0}, CHGNet \cite{Deng2023}, and M3GNET \cite{Chen2022}. For MACE we use the latest $L=2$ version, for CHGNet we use the 0.3.0 version, and for M3GNet we use the updated version trained with DIRECT sampling of the dataset \cite{M3GNet_AL}. Specific version details can be found in Table \ref{tab:models_version}.

\begin{figure}[h!]
\includegraphics[width=\linewidth]{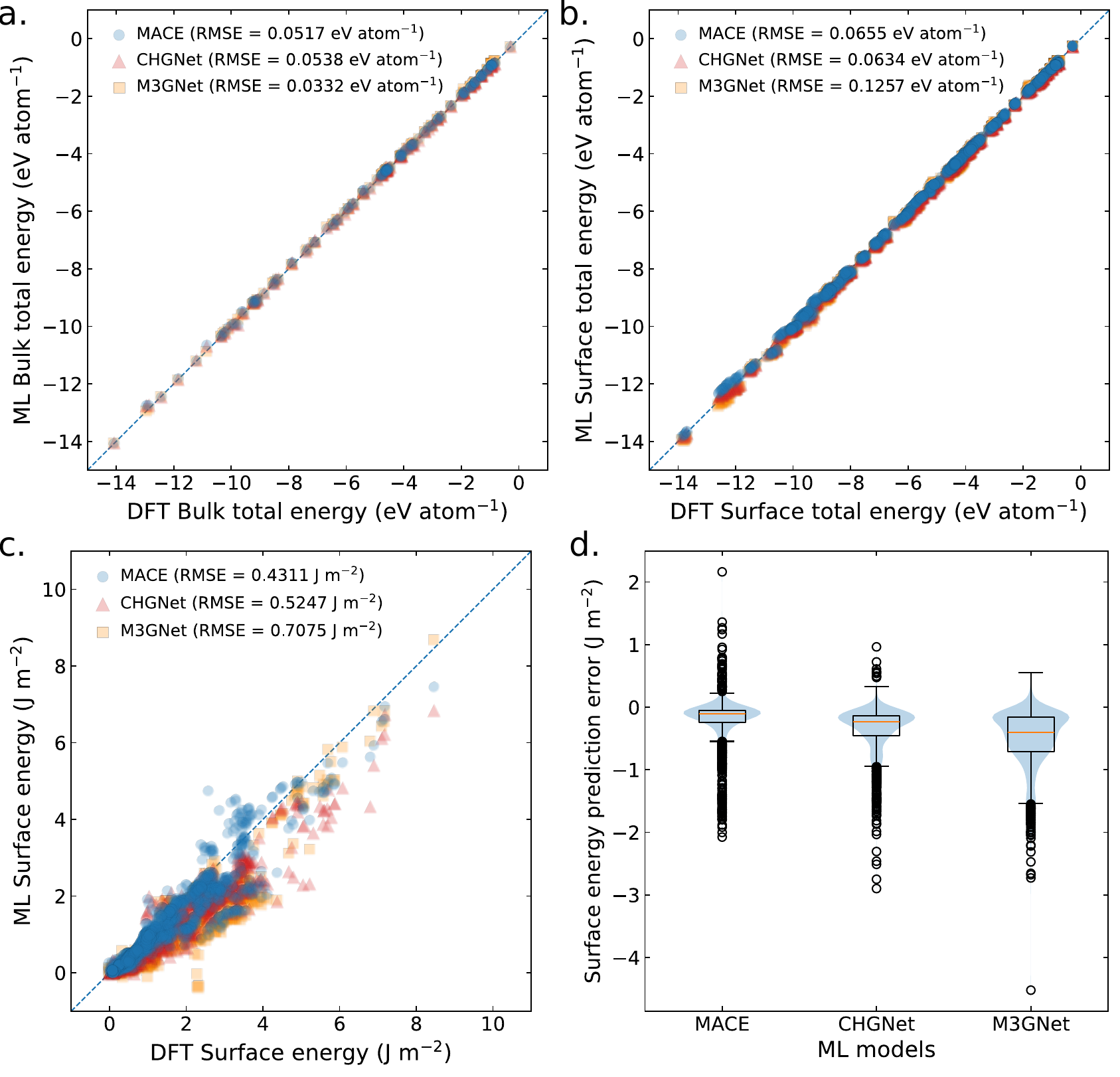}
\caption{Performance assessment of the universal interatomic potentials over the surfaces dataset. (a) Parity plot for the total energy per atom of the bulk systems that gave origin to the surfaces of the dataset. (b) Parity plot for the total energy per atom of the surfaces within the dataset. (c) Parity plot for the surface energy ($\gamma_{hkl}^\sigma$) for the surfaces within the dataset. For (a), (b), and (c), the dashed line marks the $x=y$ line. (d) Boxplot and violin plot for the error ($\gamma_{\rm ML}-\gamma_{\rm DFT}$) in the prediction of the surface energy ($\gamma_{hkl}^\sigma$) from the three universal interatomic potentials evaluated. The horizontal lines in the middle of the boxes mark the medians. The boxes are plotted from the first to the third quartile. The whiskers extend to $1.5$ times the interquartile range from the third (above) and first (below) quartile and empty circles mark the outliers (data points outside the whiskers).}
\label{fig:parity_plots}
\end{figure}

Initially, we gauge their performance in predicting the total energy of the bulk structures. Given their nature as \textit{universal} MLIPs trained on bulk data sourced from the Materials Project, a high level of accuracy is anticipated for these specific structures. Figure ~\ref{fig:parity_plots}a shows the performance of UIPs on the bulk systems that give origin to the surfaces of the dataset. As anticipated, the universal models demonstrate proficiency in predicting the total energy of bulk structures. However, their performance falls short of the precision achievable by specialized models tailored to individual systems, which in turn are expected to be below 1 meV atom$^{-1}$ for selected systems \cite{Zuo2020}. In contrast, the Matbench Discovery \cite{MatbenchDiscovery} indicates that these universal models yield RMSE values close to 100 meV atom$^{-1}$ on their test data.

Figure \ref{fig:parity_plots}b illustrates the prediction error for the total energy of surfaces. As these structures differ from the models' training set, the error rate increases. Surprisingly, CHGNet exhibits greater accuracy than both MACE and M3GNet in this context, with MACE surpassing M3GNet. This observation highlights the inherent trade-off between accuracy and generalization as M3GNet's predictive capability diminishes when extrapolating to unknown data dissimilar to the training set.

Next, we evaluate these UIPs on predicting the surface energy. The surface energy $\gamma_{hkl}^\sigma$ of a surface is defined by its Miller indexes $(\mathrm{h\,k\,l})$ and termination $\sigma$ as calculated from the following expression

\begin{equation}
    \gamma_{hkl}^\sigma = \frac{E_{\rm slab}^{hkl,\sigma} - n_{\rm slab}^{hkl,\sigma} \epsilon_{\rm bulk}}{2 A_{\rm slab}^{hkl,\sigma}}
\end{equation}

\noindent where $E_{\rm slab}^{hkl,\sigma}$ is the surface (slab) total energy, $n_{\rm slab}^{hkl,\sigma}$ is the number of sites in the surface slab, $\epsilon_{\rm bulk}$ is the bulk total energy per atom, and $A_{\rm slab}^{hkl,\sigma}$ is the area of the surface defined by the $(\mathrm{hkl})$ indexes.

Figure \ref{fig:parity_plots}c shows the parity plot for these models and Fig.~\ref{fig:parity_plots}d their error distribution. In the case of surface energy, MACE is the most accurate, however CHGNet follows in second and M3GNet in third. The three UIPs mostly underestimate the surface energy with few cases of overestimation. This behavior is reported as a ``softening'' of the potential energy surface (PES) of UIPs, also reported by Deng \textit{et al.}\cite{deng2024overcoming_softening}. Ideally, accuracy over bulk and surface total energy is desired to achieve accurate surface energies, however, error cancellation also plays a part in this analysis, therefore keeping in comparison the bulk and surface total energy is essential to gain insight into these models' performance. Additionally, we do not observe any correlation between the predictions of each model (see the Supplemental Information), that is, the errors are inherent from each model's limits due to their different training set, representations, and architecture.

\subsection{Understanding errors}
\subsubsection{System chemistry}

\begin{figure*}[ht!]
\includegraphics[width=\linewidth]{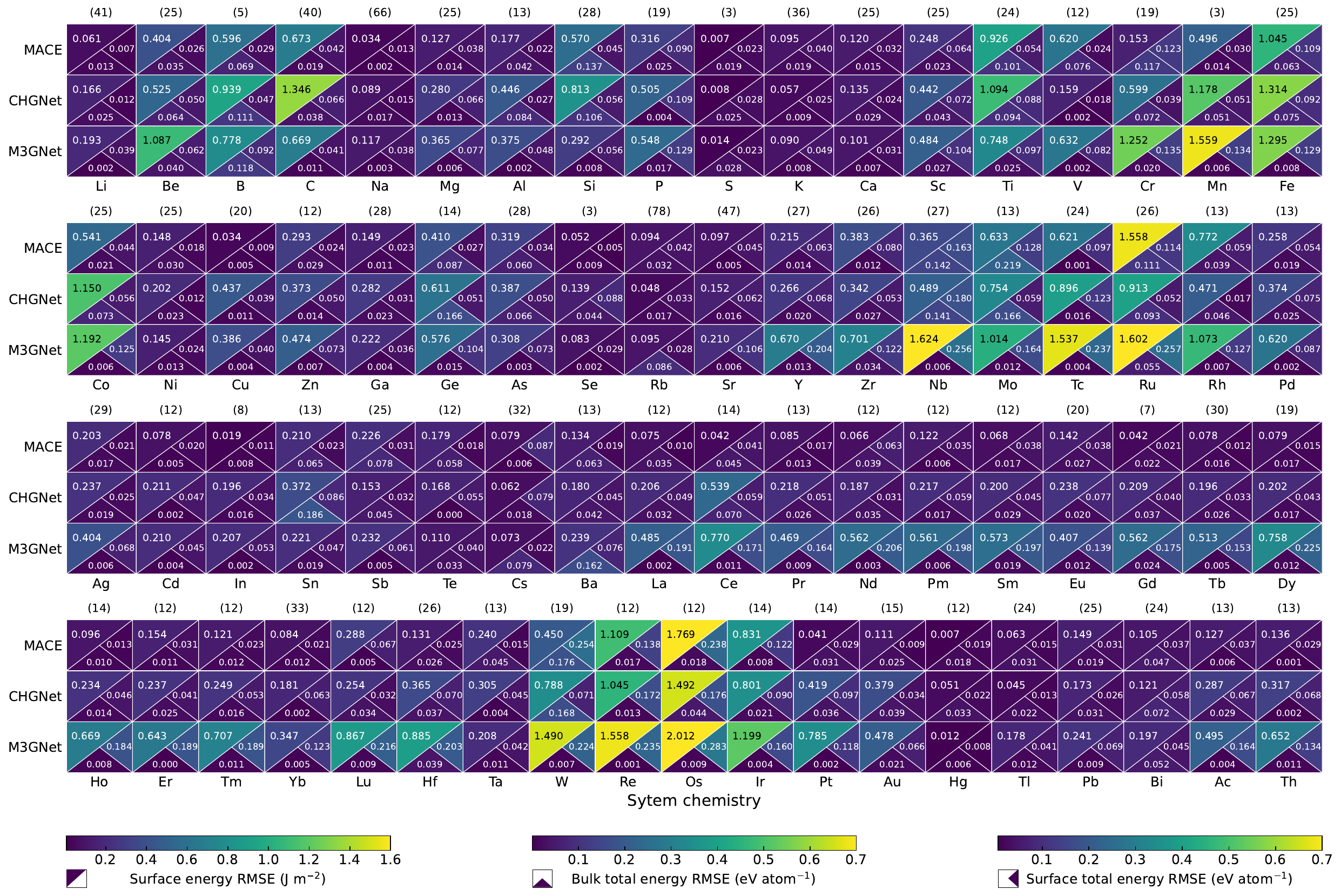}
\caption{Root mean squared error (RMSE) of the universal interatomic potentials concerning surface chemistry. The upper triangle is surface energy ($\gamma_{hkl}^\sigma$) RMSE, the lower leftmost triangle is the bulk total energy per atom RMSE and the lower rightmost triangle is the surface total energy per atom RMSE. The numbers in parenthesis are the number of surface structures evaluated for each chemical element.}
\label{fig:chemistry}
\end{figure*}

Next, we show the RMSE of these models separated by system chemistry. Figure \ref{fig:chemistry} shows a comparison between the RMSE on the surface energy and also bulk and surface total energy per atom, which we take as the limiting cases for the accuracy of each model. Figure \ref{fig:chemistry} illustrates that errors in bulk predictions are not consistently correlated with errors in surface energy predictions. For example, while MACE shows poor performance in predicting the surface energy of iron (Fe), its bulk predictions are relatively accurate. MACE demonstrates small errors for several other systems, for instance, copper (Cu) is well represented by MACE on average. Likewise, the three UIPs exhibit a higher frequency of large errors, particularly for elements such as boron (B), carbon (C), manganese (Mn), iron (Fe), niobium (Nb), molybdenum (Mo), technetium (Tc), ruthenium (Ru), rhenium (Re), osmium (Os), iridium (Ir), and platinum (Pt). Both MACE and CHGNet show fewer problematic predictions compared to M3GNet. Moreover, elements such as carbon (C), manganese (Mn), tungsten (W), iron (Fe), rhenium (Re), osmium (Os), and iridium (Ir) also require careful consideration, as these elements also present challenges for the UIPs.

The comparison with the surface total energy shows that they are the major root for the deviations of the surface energy predictions, as expected. Surfaces are not included in the training set of these universal models, therefore they struggle with extrapolating from the bulk structures. Also, since the three models share difficulties in certain elements, these errors are expected to be derived from the dataset. It is also impressive that a much smaller model M3GNet (227.5k parameters) can provide predictions as accurate as MACE and CHGNet (5.7M and 412.5k parameters, respectively) for most elemental systems.

\subsubsection{System global geometries (surfaces structures)}

\begin{figure*}[ht!]
\includegraphics[width=\linewidth]{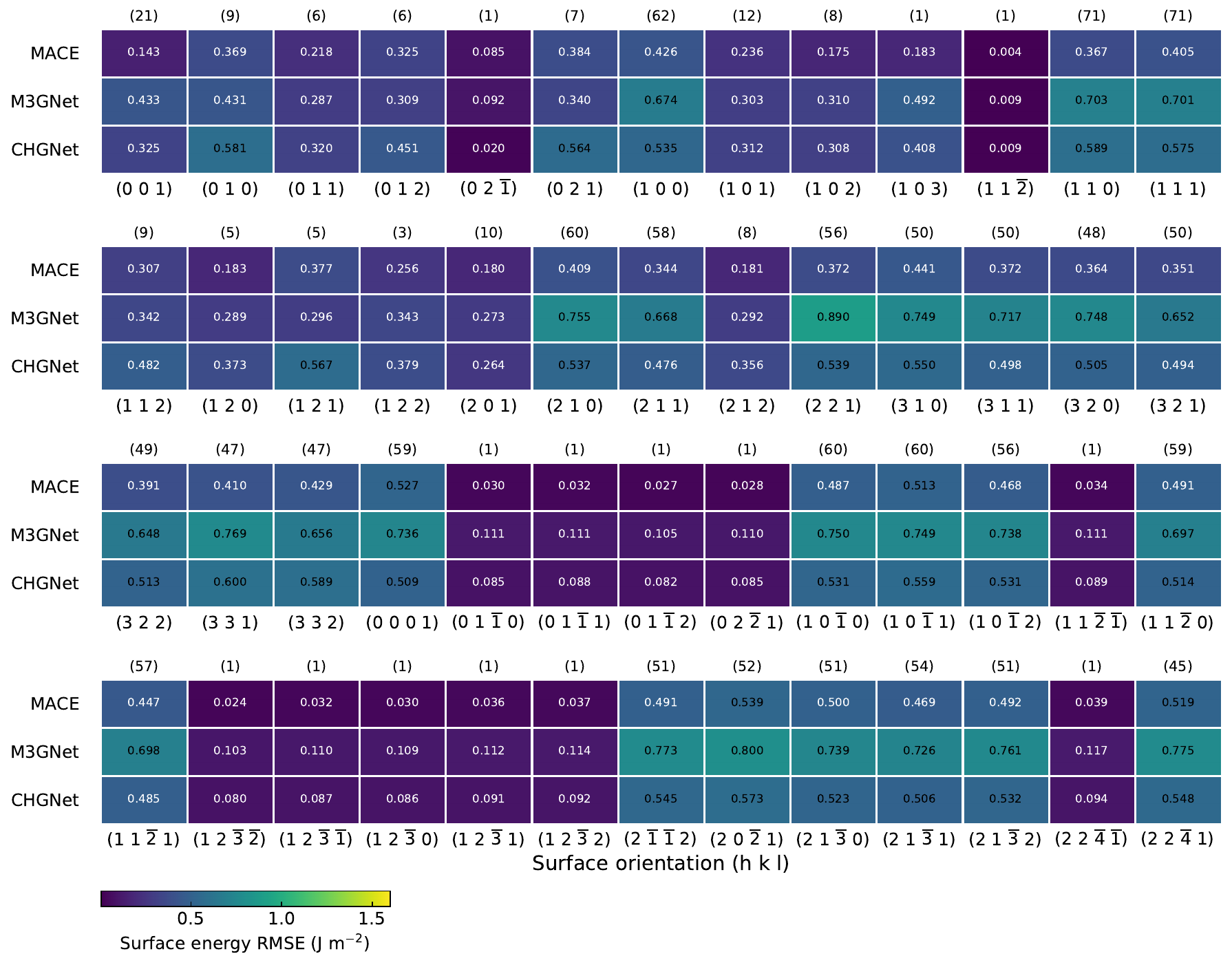}
\caption{Root mean squared error (RMSE) of the universal interatomic potentials concerning surface orientation $(h\,k\,l)$. The numbers in parenthesis are the number of surface structures evaluated for each Miller index.}
\label{fig:comparison_table_surface_miller_index}
\end{figure*}

Figure \ref{fig:comparison_table_surface_miller_index} shows the RMSE of each UIP separated by surface $(h\,k\,l)$ index. Although one expects that lower-angle surfaces show fewer errors because the atoms would keep a higher coordination degree, this depends on the termination of the slab, therefore extracting correlations with the surface orientation in such case is challenging. Across the different surface orientations, all UIPs show slightly similar errors in their prediction. Several cases of hexagonal systems appear to have a higher error than cubic ones. However, higher errors are spread across the different orientations.

\subsubsection{System local geometries (local environments, beyond surfaces structures)}

\begin{figure}[htb]
\includegraphics[width=\linewidth]{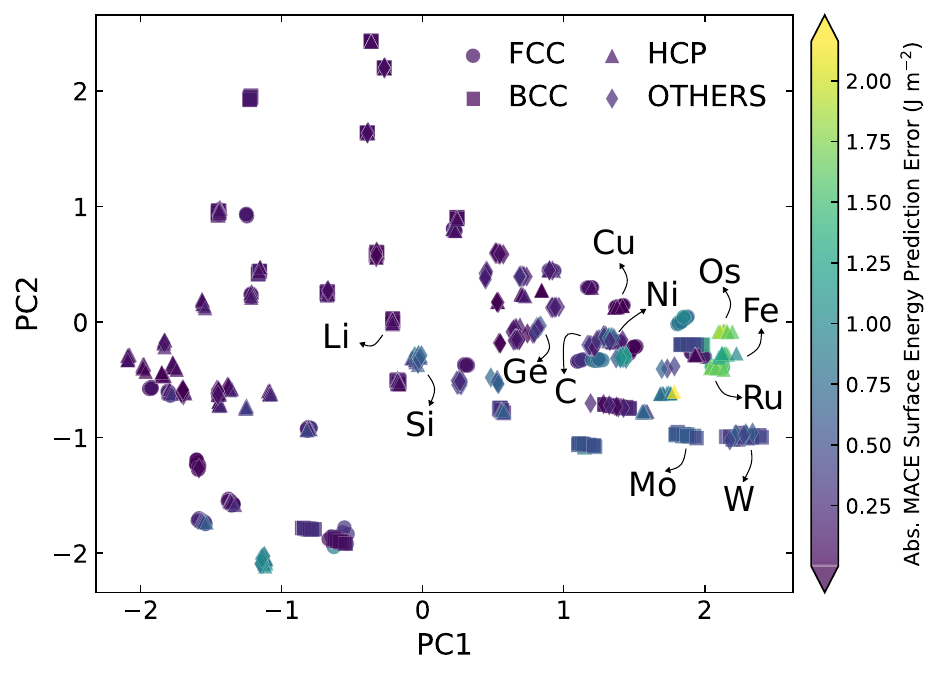}
\caption{Kernel PCA (kPCA) map of the surfaces dataset, where each point corresponds to a structure and distances represent differences in input features. kPCA map with MACE representation and REMatch kernel, colored by the absolute surface energy error given by the MACE model, with selected elements highlighted. Markers represent the lattice structure of the bulk that generated each surface, with FCC, BCC, and HCP crystalline systems highlighted.}
\label{fig:surface_descr_map}
\end{figure}

Figure \ref{fig:surface_descr_map} shows a kernel PCA (kPCA) map created to represent each structure and its chemistry accounting for their local environments. The descriptor used for each structure is created using MACE's representation with the same hyperparameters used for training the universal model, that is, a radius cutoff of $6.0\;\text{\AA}$ and three layers of 128 channels with a maximum angular momentum of $l_{\rm max}=2$. For each structure, we use a simple average between the atomic sites, and the comparison between structures is done by the REMatch kernel \cite{Sandip2016_ReMatchKernel}.

 Figure \ref{fig:surface_descr_map} shows the kPCA map colored by the absolute error given by the MACE model for surface energy. Interestingly, the MACE representation is enough to single out most systems. The representation seems to take great advantage of the different chemical species within the dataset. As shown in Fig.~\ref{fig:chemistry}, osmium (Os) presents most of the error from MACE. Additionally, in Fig.~\ref{fig:surface_descr_map} we also try to associate the kPCA map with the lattice of each original bulk structure. We highlight the surfaces that originate from face-centered cubic (FCC), body-centered cubic (BCC), and hexagonal closed-packed (HCP) lattices. The error is not directly dependent on the lattice symmetry. Detailed analysis of each system chemistry on the kPCA map is available in the Supporting Information.

\subsection{Comparison: universal vs specialized MLIPs}

\subsubsection{Universal vs NequIP/MTP}

Our next step in this assessment is to address the question: ``If one is interested in materials' surfaces, are universal interatomic potentials a good starting point, or should one train a specialized interatomic potential for the task at hand?''. For this comparison, we started by selecting candidates for specialized MLIPs. Keeping in mind the performance-computational cost trade-off, the moment tensor potential (MTP) \cite{Shapeev2016_MTP} formalism is ideal since it can attain very low test errors with a rather small dataset and number of parameters \cite{Zuo2020,Xie2023}. \citet{Zuo2020} performed a systematic benchmark of ``pre-universal'' MLIPs and made available not only the training and test sets of their performance and computational cost assessment, but also the models they have trained for the elements: nickel (Ni), copper (Cu), lithium (Li), molybdenum (Mo), silicon (Si), and germanium (Ge). We called MTP [Ong21] the set of MTP models as available from ref. \cite{Zuo2020} for each of the elements considered in their work.

\begin{figure}[htb!]
\includegraphics[width=\linewidth]{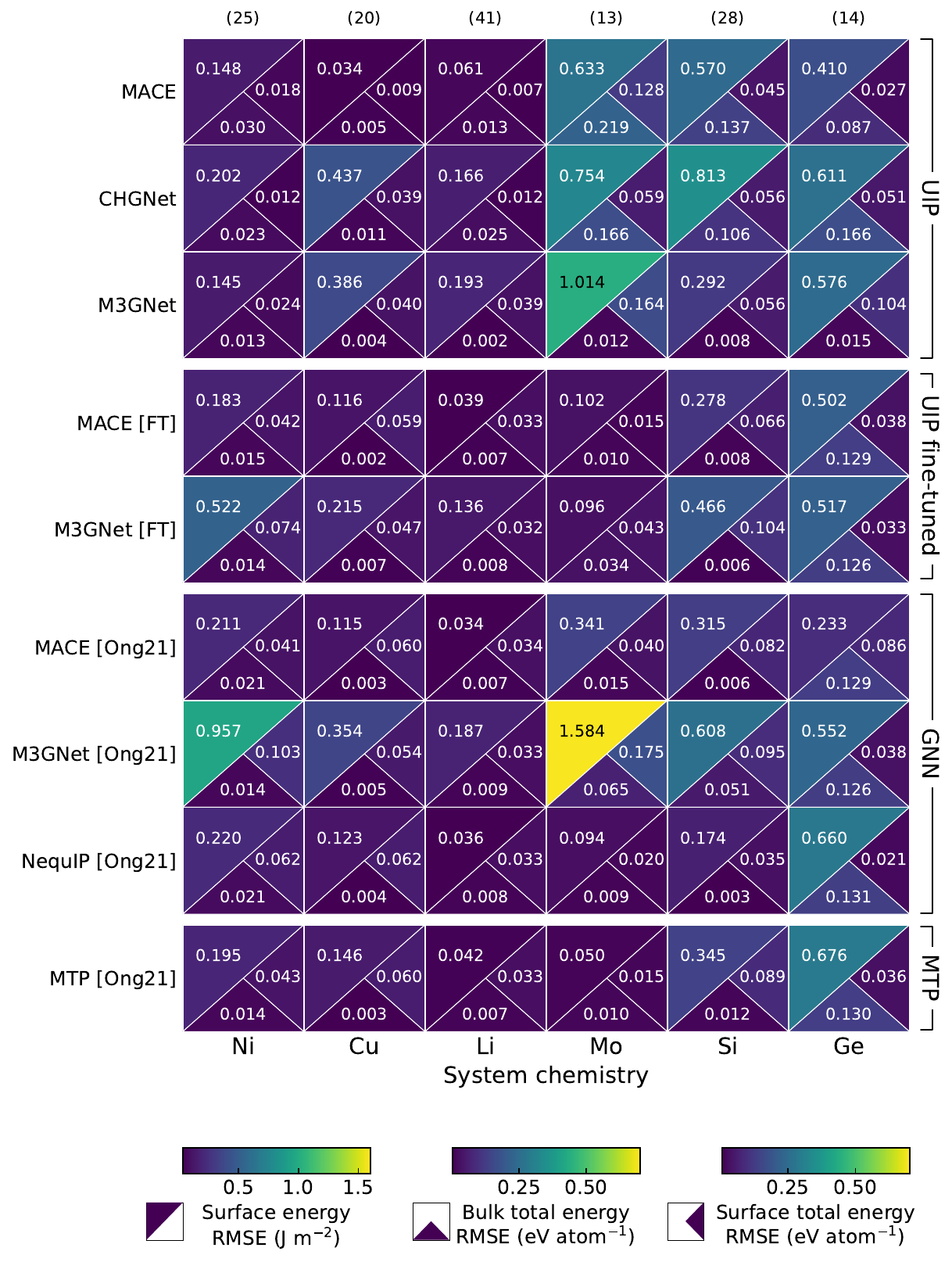}
\caption{Performance comparison between universal, specialized, and universal fine-tuned interatomic potentials. The root mean squared errors (RMSE) of the universal interatomic potentials are compared to the surface chemistry, for selected elements. The upper triangle is the surface energy ($\gamma_{hkl}^\sigma$) RMSE, the lower leftmost triangle is the bulk total energy per atom RMSE and the lower rightmost triangle is the surface total energy per atom RMSE. The numbers in parenthesis are the number of surface structures evaluated for each chemical element.}
\label{fig:comparison_table}
\end{figure}

We added the Neural Equivariant Interatomic Potential (NequIP) to our comparison as they are equivariant neural networks, similar to the UIPs, and highly efficient in data use. We trained a NequIP model using ref. \cite{Zuo2020} training and test sets, which we called NequIP [Ong21]. Our comparison contrasts the UIPs: MACE, M3GNet, and CHGNet with two sets of specialized MLIPs: MTP [Ong21], and NequIP [Ong21], as shown in Fig.~\ref{fig:comparison_table} and summarized in Table \ref{tab:rmse_comparison}. See the Supporting Information for the parity plot containing all these models for this reduced set of elements.

First, by comparing the RMSE in Table \ref{tab:rmse_comparison} with the RMSE of the full set of elements from the MPtrj dataset, the RMSE of UIPs shows a different trend. For this smaller set of elements, although MACE and CHGNet are still not the most accurate total energy prediction on the bulk structures, there is not a significant increase in the error for surface total energy predictions, in fact, the RMSE decreases for the prediction of surface total energy. M3GNet on the other hand shows an error increase for the surfaces. For surface energy per area, MACE shows the smallest error among the UIPs.

We will now compare the MTP [Ong21] and NequIP [Ong21] models with UIPs. These models trained with the [Ong21] dataset present an overall better performance. We attribute this to the sheer influence of the construction of the dataset. The [Ong21] dataset is more diverse in types of structures, containing: MD trajectory snapshots, defects, deformations, and even a few surfaces, while the MPtrj, although very large datasets contains only bulk structures close to their relaxed geometries and relaxation trajectories. Comparison between MTP and NequIP brings the evident power of MTP. Although NequIP showcases smaller RMSE, MTP is not far behind, especially for bulk structures. However, as we go to surfaces, NequIP showcases higher generalization power. Comparison between specialized and universal IPs demonstrates that UIPs could have their generalization power improved to yield predictions for surfaces with sufficient accuracy.

\begin{table}[h!]
\centering
\caption{Root mean squared errors (RMSE) for universal interatomic potentials, specialized machine learning interatomic potentials, and fine-tuned universal interatomic potential for a selected set of elements: Ni, Cu, Li, Mo, Si, and Ge. For the column type, we label the models as universal interatomic potential (UIP), UIP fine-tuned (UIP-FT), graph neural network (GNN), and moment tensor potentials (MTP).}
\label{tab:rmse_comparison}
\resizebox{\linewidth}{!}{%
\begin{tabular}{rcccc}
\hline\hline
\multicolumn{1}{c}{}         & \multicolumn{3}{c}{RMSE over structures with Ni, Cu, Li, Mo, Si, and Ge}                                                                                                                                                              &  \\
\multicolumn{1}{c}{ML model} & \begin{tabular}[c]{@{}c@{}}Bulk total energy\\ (eV atom$^{-1}$)\end{tabular} & \begin{tabular}[c]{@{}c@{}}Surface total energy\\ (eV atom$^{-1}$)\end{tabular} & \begin{tabular}[c]{@{}c@{}}Surface energy\\ (J m$^{-2}$)\end{tabular}& Type \\ \hline
MACE                         & 0.0856                                                                       & 0.0455                                                                          & 0.3512                                                               & UIP \\
CHGNet                       & 0.0791                                                                       & 0.0386                                                                          & 0.5128                                                               & UIP \\
M3GNet                       & \textbf{0.0084}                                                              & 0.0704                                                                          & 0.4246                                                               & UIP \\  \hline
MACE [FT]                    & 0.0464                                                                       & \textbf{0.0139}                                                                 & \textbf{0.2228}                                                               & UIP-FT \\
M3GNet [FT]                  & 0.0399                                                                       & 0.0590                                                                          & 0.3852                                                               & UIP-FT \\ \hline
MACE [Ong21]                 & 0.0372                                                                       & 0.0580                                                                          & \textbf{0.2144}                                                      & GNN \\
M3GNet [Ong21]               & 0.0448                                                                       & 0.0859                                                                          & 0.7247                                                              & GNN \\
NequIP [Ong21]               & 0.0378                                                                       & 0.0542                                                                          & 0.2696                                                               & GNN \\
MTP [Ong21]                  & 0.0371                                                                       & 0.0537                                                                          & 0.2820                                                               & MTP \\ \hline\hline
\end{tabular}%
}
\end{table}

\begin{figure*}
\includegraphics[width=\linewidth]{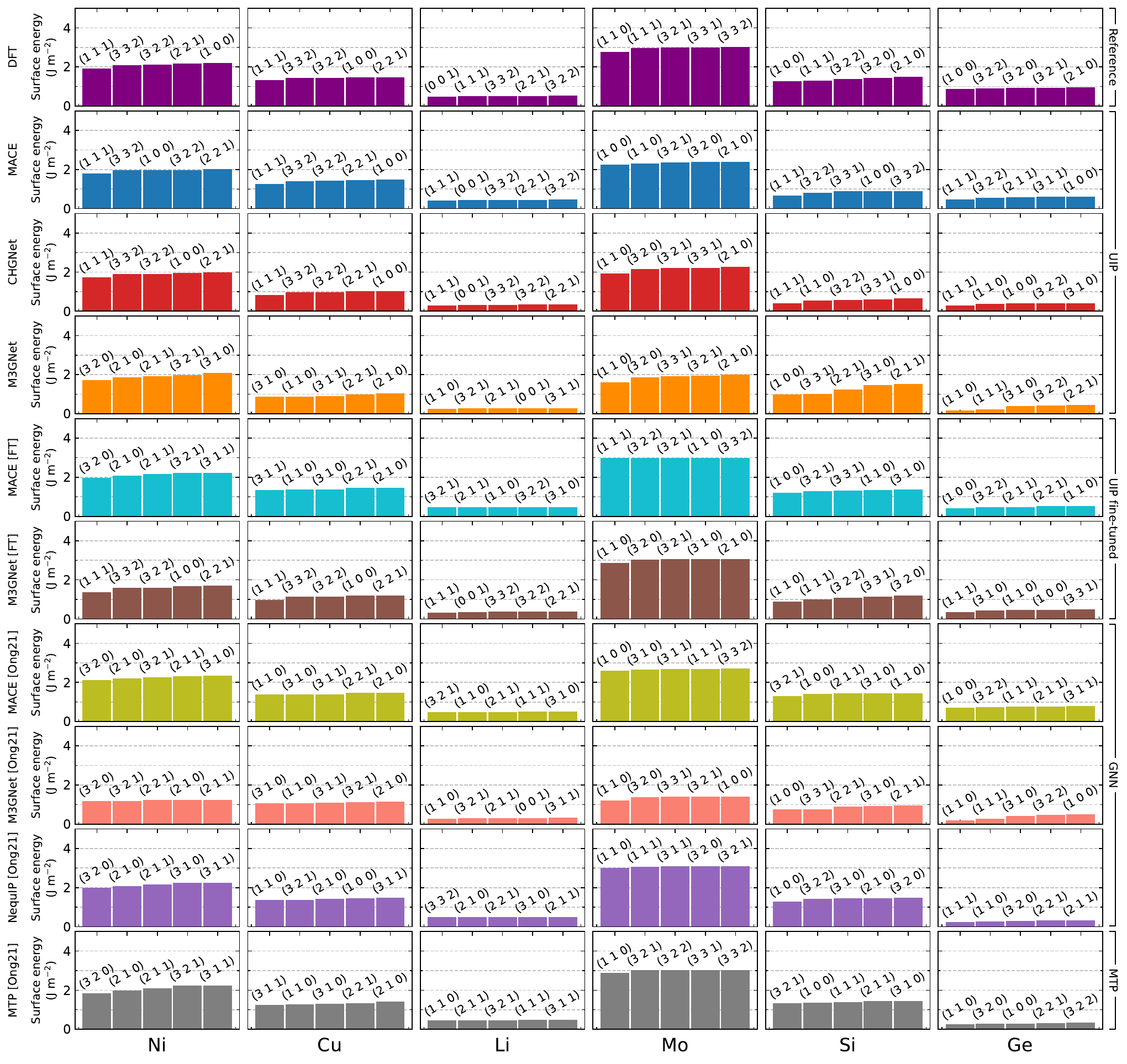}
\caption{Performance comparison between universal, specialized, and universal fine-tuned interatomic potentials for prediction of relative surface energy order. The evaluated surfaces are derived from the bulk structures \texttt{mp-23}, \texttt{mp-30}, \texttt{mp-51}, \texttt{mp-129}, \texttt{mp-149}, and \texttt{mp-32}, for Ni, Cu, Li, Mo, Si, and Ge, respectively.}
\label{fig:surface_energy_properties}
\end{figure*}

As demonstrated in Fig.~\ref{fig:comparison_table}, for most elements where UIPs show an increased error, MTP [Ong21] and NequIP [Ong21] demonstrate high-accuracy predictions. Predictions for systems such as Mo and Si show great improvement from the specialized MLIPs. Systems containing Ge on the other hand show worse performance for bulk structures while slightly improved performance for surface structures, as expected from the results from \citet{Zuo2020}.

To understand the relevance of the dataset when comparing results between MTP and NequIP trained with [Ong21] dataset, we create another comparison, as detailed in Supporting Information, with the so-called MPtrj$^{*}$. The MPtrj$^{*}$ dataset is created by filtering the MPtrj dataset for unary structures containing the following elements: Ni (96 structures), Cu (55 structures), Li (83 structures), Mo (25 structures), Si (559 structures), and Ge (225 structures). A single MTP and NequIP model is fitted for each of these elements and compared to the [Ong21] dataset. As expected, since this new dataset has a much less diverse set of structures, with forces and stresses close to zero, these non-universal models struggle to predict accurate surface (total) energies, particularly MTP. The full comparison with the MPtrj$^{*}$ dataset and further details can be found in the Supporting Information.

\subsubsection{Universal (zero-shot) vs Universal (fine-tuned) vs From scratch models}

Since UIPs seem to not be great at dealing with surfaces, we now address the question: ``What can we do to improve them?''. One route to improving general models such as the UIPs is fine-tuning, which consists of continuing the training of these models however with a different dataset, tailored to make these universal models more specialized in particular downstream tasks. In our case, we can use the training and test sets of ref. \cite{Zuo2020} to fine-tune these UIPs. By introducing a more diverse set of structures for each chemical element, we expect that the UIP will improve in predicting surface (total) energies while taking advantage of what it has already learned from the MPtrj dataset. 
We fine-tune the MACE and M3GNet models, given the availability of the corresponding implementation. We do not fine-tune the CHGNet model since the dataset does not contain atomic magnetic moments. 
We fine-tune the MACE model for another 200 epochs using the training set from ref. \cite{Zuo2020}. The fine-tuned model is called MACE [FT], as shown in the last line of Fig. \ref{fig:comparison_table} and Table \ref{tab:rmse_comparison}. 
We include in our comparison MACE and M3GNet models trained from scratch with the same data used for fine-tuning, so-called MACE/M3GNet [Ong21]. A comparison between [FT] and [Ong21] may reveal how the model can use its previous knowledge to improve its prediction versus learning from the data from scratch.

After fine-tuning, we observe that MACE prediction of bulk total energies largely improved for most systems. It is noticeable how fine-tuning should be performed with care. As discussed before, both MTP and NequIP did not show great performance over bulk structures composed of germanium (Ge). After fine-tuning, MACE [FT] decreased its performance on germanium (Ge) bulk structures.%
The performance on surface total energies improved for molybdenum (Mo), while for the other elements it mostly decreased. For Li, Mo, and Si, we observe significant improvement over surface energies. Overall, the fine-tuned universal model represents an important improvement: for the selected systems, the fine-tuned MACE model is of similar performance to specialized MLIPs. 

On comparison between MACE [FT] and MACE [Ong21], for Ni, Cu, and Li, their accuracy is very similar, while for Mo and Si, MACE [FT] shows significantly more accurate prediction, with the opposite for Ge. A closer look at Ge reveals that performance over bulk structures is of a similar error, however, the FT model outperforms MACE [Ong21] on surface total energies, while for surface energies, MACE [Ong21] outperforms MACE [FT].

The comparison between M3GNet and M3GNet [FT] shows smaller differences, while errors on the bulk total energies increase, there are improvements on both surface total energies and surface energies, although smaller than 20\% on RMSE. On the other hand, when comparing M3GNet with M3GNet [Ong21], the benefits of the universal pretraining strategy are clearer for this model, whereas the from-scratch model shows poor performance.

\subsubsection{Surface relative energy order}

In addition to the RMSE for the surfaces, we investigate if MLIPs, especially UIPs, are able to accurately reproduce the relative energy order of surfaces with different orientations. This is particularly important when using UIPs for an initial screening of low-energy surfaces and structures, as well as to reproduce fundamental properties of materials, such as the Wulff shape.

Figure \ref{fig:surface_energy_properties} shows the 5 lowest energy surfaces found by each MLIP in comparison with our reference data, which are the DFT surface energies. The comparison uses only the surfaces derived from the most stable bulk structure of each element, specifically \verb|mp-23|, \verb|mp-30|, \verb|mp-51|, \verb|mp-129|, \verb|mp-149|, and \verb|mp-32|, for Ni, Cu, Li, Mo, Si, and Ge, respectively.

Across these six elements, MACE and CHGNet tend to reproduce well the first two surfaces for Ni and Cu. All three UIPs fail to reproduce the order of the lowest surface energy for Li, Mo, and Ge. M3GNet is able to correctly predict the lowest surface energy for Si, while MACE and CHGNet fail. In fine-tuned models, MACE [FT] reproduces the lowest energy surface for Si and Ge, and M3GNet can correctly predict the lowest surface energy for Ni, Cu, and Mo. Specialized models trained with the [Ong21] dataset also struggle. For instance, M3GNet [Ong21], NequIP [Ong21], and MTP [Ong21] can correctly predict the lowest surface energy for Mo, while M3GNet [Ong21] and NequIP [Ong21] can do so for Si, and MACE [Ong21] only for Ge.

Overall, Figure \ref{fig:surface_energy_properties} reveals a major challenge: none of the MLIPs tested is accurate enough to reproduce the order of the lowest energy surfaces. In most cases, the energy difference between the set of lowest energy surfaces is in the order of $0.01\text{--}0.1 \rm J\;m^{-2}$, while their average error for the MLIPs tested is in the order of $0.2\text{--}0.7 \rm J\;m^{-2}$.

In addition to the surface properties, we performed molecular dynamics (MD) runs for the higher RMSE surfaces of each element. To assess the stability of the trajectories from the MLIPs considered in this work, we carried out NVT MD simulations at $T = 0.8 \times \text{melting point}$ for the higher RMSE surfaces of each corresponding element for $30\;\rm ps$. For all UIPs and MACE [FT], the dynamics were stable with no qualitative difference between the MLIPs.

\section{Conclusions}

We have assessed one of the main questions for working with MLIPs: given a new problem, what is the best and most efficient way to tackle it? The answer follows at least two steps, evaluating if MLIPs are accurate enough and if not, if it is better to train a specialized MLIP from scratch or to fine-tune a universal foundational model.
From our results for surface energies, we see that the total energies for surface geometries are modestly accurate, however, not good enough for specific properties. 
Therefore, as expected, the most important aspect is the coverage of the training dataset to the target space, composed of both chemistry and structure.
The investigation of what are the most efficient strategies to cover this space is an important question for future studies, as UIPs can display low errors in structures close to equilibrium \cite{Rignanese2024}, suggesting a connection between out-of-domain distance and prediction errors \cite{Li2023,deng2024overcoming_softening}.

In terms of efficiency, fine-tuning UIPs can greatly accelerate the training by incorporating alchemical transferable knowledge from the large bulk dataset (in this case, MPtrj), thus requiring only a modest dataset to achieve sufficient accuracy in specialized tasks.
The combination of active learning strategies using cheap UIP trajectories to generate synthetic datasets for pre-training MLIPs can also greatly increase the efficiency of generating new datasets and expanding current ones \cite{M3GNet_AL,Gardner2024}.
Making use of already existing datasets, although at different fidelity levels, can be explored in multi-fidelity or multi-modality approaches, recently shown to be beneficial and accelerating fine-tuning more than 100$\times$ \cite{Wang2023_dpa2}.
Additionally, there is still plenty of exploration of potential gains related to the architectural design of the models. Beyond graph Atomic Cluster Expansions \cite{Drautz2024}, models based on transformers such as EquiformerV2 \cite{Smidt2024_equiformer2} display improved data efficiency, and MatterSim models \cite{Lu2024_mattersim} using this architecture are also comparatively more accurate.

The performance of MLIPs for surfaces is important because, in materials science, a plethora of different phenomena take place on the surface of materials. Examples are adsorption and absorption of molecules and gases, catalysis, electrochemical processes such as water splitting, deposition and growth of new materials and phases such as heterostructures and nanoparticles. The generalization including both extensivity and transferability of MLIPs allows for simulations and understanding of ever larger and complex systems at an efficient computational cost. This enables the study of nanoscale and quantum materials systems such as 2D materials and their twisted interfaces, as well as more general global challenges involving molecules and materials, such as in energy, health, and the environment \cite{Focassio2022,Fiuza2024,Mortazavi2021,ML2D_2019,Tritsaris2021,Li2023_2D,Schleder2023,Hsieh2023,Mortazavi2021_continuum,Verissimo2024,Petry2021,Nascimento2022}.
As these models become more used, understanding their application limits and broadening them is of great importance to accelerate materials science research and applications of these models to different systems and use cases.

We envision that a promising step forward for the community is to create a universal training dataset for models to be trained upon, including not only all elements, but more importantly, intelligent sampling of the full combinatorial materials space composed of molecules, bulk solids, surfaces, interfaces, defects, multi-element alloys, and so on. Even though the currently available materials databases are a great starting point, they have been created with specific tasks in mind, focused on materials discovery. Compared to the huge size of the materials space, they still represent a very scarce fraction of all possible combinations. From a broader perspective, incorporating additional properties such as magnetic moments \cite{Deng2023} or charge densities \cite{Shen2022,Focassio2023}, can be promising routes for further advancements \cite{Ko2023}. There are still plenty of advances ahead. 

\section{Computational Details}

\subsection{Datasets}

The Materials Project surface dataset (crystalium) \cite{Tran2016} was accessed on Feb. 2023 through Materials Project's API \cite{Jain2013_MP}.

We call MPtrj$^{*}$ the MP data that is filtered for unary structure from the following set of elements from the MPtrj \cite{Deng2023_mptrj} dataset: Ni (96 structures), Cu (55 structures), Li (83 structures), Mo (25 structures), Si (559 structures), Ge (225 structures). Accordingly, for MTP and NequIP, a different model is fit for each element. The entire MPtrj$^{*}$ dataset is used for training and the test is performed on the MP surfaces dataset. Further details can be found in the Supporting Information.

The so-called [Ong21] dataset is made available from ref. \cite{Zuo2020} and contains structures for Ni (263/31 structures), Cu (262/31 structures), Li (241/29 structures), Mo (194/23 structures), Si (214/25 structures), Ge (228/25 structures) for training/testing tasks. Further details can be found on ref. \cite{Zuo2020} and the \verb|maml| package on \url{https://github.com/materialsvirtuallab/maml}.

\subsection{Universal potentials}

\subsubsection{Model's versions}

Table \ref{tab:models_version} shows the details and versions of the UIPs tested in this work.

\begin{table}[!h]
    \centering
    \caption{Details and version of UIPs tested.}
    \label{tab:models_version}
    \resizebox{\linewidth}{!}{%
    \begin{tabular}{rlccc}
    \hline\hline
        \multicolumn{1}{c}{Model} & \multicolumn{1}{c}{Version} & Model size & Dataset & Data size  \\ \hline
         M3GNet \cite{Chen2022}      & 2021.2.8-DIRECT \cite{M3GNet_AL}   & 1.1M & MPF \cite{Chen2022}         & 185.6k \\
         CHGNet \cite{Deng2023}      & v0.3.0             & 412.5k & MPtrj \cite{Deng2023_mptrj} & 1.58M \\
         MACE \cite{Batatia2022mace} & 2024.01.07-128-L2  & 5.7M   & MPtrj \cite{Deng2023_mptrj} & 1.58M \\ \hline\hline
    \end{tabular}
    }
\end{table}

\subsubsection{Fine-tuning}

We fine-tune the MACE model with the aid of the training scripts provided with the \verb|mace| code package available from \url{https://github.com/ACEsuit/mace}. The initial model for fine-tuning MACE is the MACE-MP-0 (for $L=2$) model, as specified in Table \ref{tab:models_version}, used throughout the text as MACE, available from \url{https://huggingface.co/cyrusyc/mace-universal} (2024-01-07-mace-128-L2\_epoch-199.model). The data used for fine-tuning is available from ref. \cite{Zuo2020}. Fine-tuning was performed for a maximum of 200 epochs, training was stopped if the validation metric did not improve for 50 epochs. Further details, as well as the trained model, are made available as Supporting Information.

Fine-tuning the M3GNet model is performed with aid of the \verb|matgl| code package available from \url{https://github.com/materialsvirtuallab/matgl}. The initial model for fine-tuning M3GNET is the M3GNET-DIRECT model trained in ref. \cite{M3GNet_AL}, as specified in Table \ref{tab:models_version}, and used throughout the text as M3GNET. The data used for fine-tuning is available from ref. \cite{Zuo2020}. Fine-tuning was performed for a maximum of 200 epochs, training was stopped if the validation metric did not improve for 50 epochs. Further details, as well as the trained model, are made available as Supporting Information.

Fine-tuning for both MACE and M3GNet was performed using a single NVIDIA Tesla P100 16 GB GPU, taking 14 hours on average for MACE and 3 hours on average for M3GNet using 1402 structures in both cases.

\subsection{Specialized potentials}

\subsubsection{MACE from scratch}

We trained a MACE model using the training scripts provided with the \verb|mace| code package available from \url{https://github.com/ACEsuit/mace}. Hyperparameters for training MACE were set to spherical expansion of up to $l_{\rm max} = 3$, and 4-body messages in each layer; 128-channel dimension for tensor decomposition; radius cutoff of 6 \AA;  10 Bessel functions for radial basis; three hidden layers of 64 hidden units; and $L=2$ resulting in the \texttt{128x0e+128x1o+128x2e} irreducible representations. The weights for the Huber loss considering energy, forces, and stress were set to $\lambda_E=1$, $\lambda_F=10$, and $\lambda_\sigma=100$, following ref. \cite{MACE_MP-0}. All other hyperparameters were the same as in ref. \cite{MACE_MP-0}. The data used for training is available from ref. \cite{Zuo2020}. The trained models are available as Supporting Information.

\subsubsection{M3GNet from scratch}

We trained a M3GNet model using the training scripts provided with the \verb|matgl| code package available from \url{https://github.com/materialsvirtuallab/matgl}. Hyperparameters for training M3GNet were set to a 1:1:0.1 weight ratio for energy (eV atom$^{-1}$), force (eV \AA$^{-1}$), and stress (GPa) in a Huber loss function with $\delta = 0.01$, an Adam optimizer with initial rate of $10^{-3}$ and a cosine decay to 1\% of the original value in 100 epochs. The training was performed for 200 epochs with 50 epochs of patience over the validation total loss. The number of hidden units, dimension of node, and edge embeddings were set to 128. The radius cutoff was set to $5$ \AA, and all other hyperparameters were set to be the same as in M3GNet trained in ref. \cite{M3GNet_AL}The data used for training is available from ref. \cite{Zuo2020}. The trained models are available as Supporting Information.

\subsubsection{MTP models}

The moment tensor potential (MTP) models construct a contracted rotationally invariant representation of the atomic local environments with tensors from a set of basis functions, these are used to build a linear expansion of the potential energy as a function of the atomic representation \cite{Shapeev2016_MTP}. The so-called MTP [Ong21] models are made available from ref. \cite{Zuo2020} through the \verb|maml| package on \url{https://github.com/materialsvirtuallab/maml}. The MTP models trained with the MPtrj$^{*}$ dataset are trained using the set of hyperparameters selected via grid search. Further details and the trained models are available as Supporting Information.

\subsubsection{NequIP models}

Neural Equivariant Interatomic Potential (NequIP) is a highly accurate approach to generate MLIP models with outstanding data efficiency. The architecture of the model yields an E(3)-equivariant graph neural network in which the features are updated in a message passing scheme and the rotation equivariance is achieved by convolution filters via tensor product of radial functions and spherical harmonics \cite{Batzner2022,e3nn}. In this case, the convolution operation compromises the atoms close to a central atom within a cutoff radius which results in semi-local environments. Herein we construct NequIP models from MPtrj$^{*}$ and [Ong21] datasets. The model is trained using the \verb|nequip| package available from \url{https://github.com/mir-group/nequip}. The set of hyperparameters is selected via grid search. Further details and the trained models are available as Supporting Information.

\section*{Supporting Information}

Dataset exploration: surface energy data distribution;
Evaluation of UIPs: correlation between UIPs' predictions, understanding the errors by system chemistry;
Comparison between UIPs, specialized MLIPs, and fine-tuned UIP: parity plot and root mean squared errors table of all models;
Performance and computational cost concerning fine-tuning training volume;
Hyperparameter optimization of specialized IPs.
All trained models and code are available as files with instructions on their use in the following repository \cite{zenodo}: \url{https://doi.org/10.5281/zenodo.11391989}.

\section*{Acknowledgements}
The authors acknowledge support from CNPq project no. 422069/2023-0 and CNPq - INCT (National Institute of Science and Technology on Materials Informatics), grants no. 371610/2023-0. LPMF is a CNPq scholarship holder - Brazil (150597/2023-1).

%

\clearpage


\section*{Supplementary material}

## III. COMPARISON BETWEEN UIPS, SPECIALIZED MLIPS, AND FINE-TUNED UIP

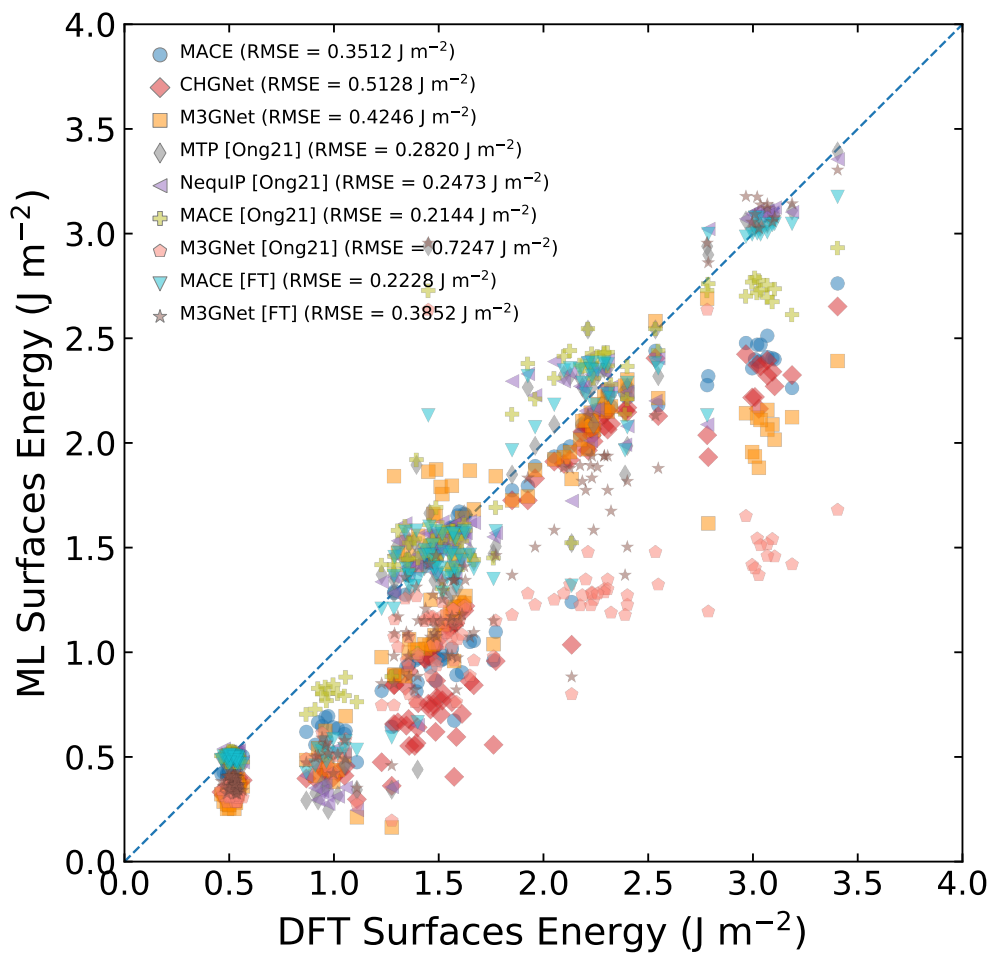

**Figure S4**. Performance assessment of the universal interatomic potentials over the surfaces dataset. Parity plot for the surface energy ($\gamma_{hkl}^\sigma$) for the surfaces within the dataset, including all models evaluated, both universal and specialized IPs. The dashed line marks the ideal $x = y$ line.

### A. MPtrj* dataset

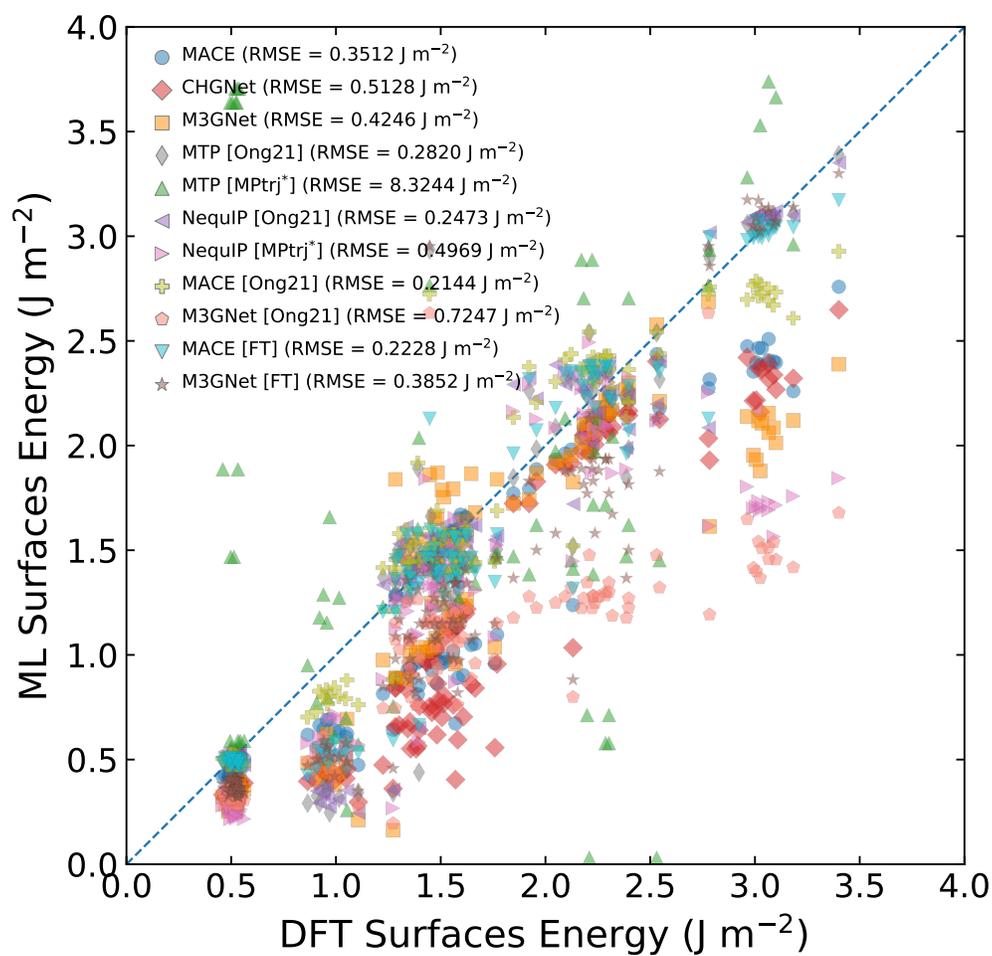

**Figure S5**. Performance assessment of the universal interatomic potentials over the surfaces dataset. Parity plot for the surface energy ($\gamma_{hkl}^{\sigma}$) for the surfaces within the dataset, including all models evaluated, both universal and specialized IPs. The dashed line marks the ideal $x = y$ line.

**Table S1**. Root mean squared errors (RMSE) for universal interatomic potentials, specialized machine learning interatomic potentials, and fine-tuned universal interatomic potential for a selected set of elements: Ni, Cu, Li, Mo, Si, and Ge. For the column type, we label the models as universal interatomic potential (UIP), local representation interatomic potential (LRIP), graph neural network (GNN), and UIP fine-tuned (UIP-FT).

| ML model | RMSE over structures with Ni, Cu, Li, Mo, Si, and Ge | | | |
| | Bulk total energy (eV atom$^{-1}$) | Surface total energy (eV atom$^{-1}$) | Surface energy (J m$^{-2}$) | Type |
| --- | --- | --- | --- | --- |
| MACE | 0.0856 | 0.0455 | 0.3512 | UIP |
| CHGNet | 0.0791 | 0.0386 | 0.5128 | UIP |
| M3GNet | **0.0084** | 0.0704 | 0.4246 | UIP |
| MACE [FT] | 0.0464 | **0.0139** | **0.2228** | UIP-FT |
| M3GNet [FT] | 0.0399 | 0.0590 | 0.3852 | UIP-FT |
| MACE [Ong21] | 0.0372 | 0.0580 | **0.2144** | GNN |
| M3GNet [Ong21] | 0.0448 | 0.0859 | 0.7247 | GNN |
| NequIP [Ong21] | 0.0378 | 0.0542 | 0.2696 | GNN |
| MTP [Ong21] | 0.0371 | 0.0537 | 0.2820 | LRIP |
| MTP [MPtrj*] | 0.0187 | 0.9370 | 8.3244 | LRIP |
| NequIP [MPtrj*] | 0.0182 | 0.0880 | 0.5137 | GNN |

## IV. PERFORMANCE AND COMPUTATIONAL COST CONCERNING FINE-TUNING TRAINING VOLUME

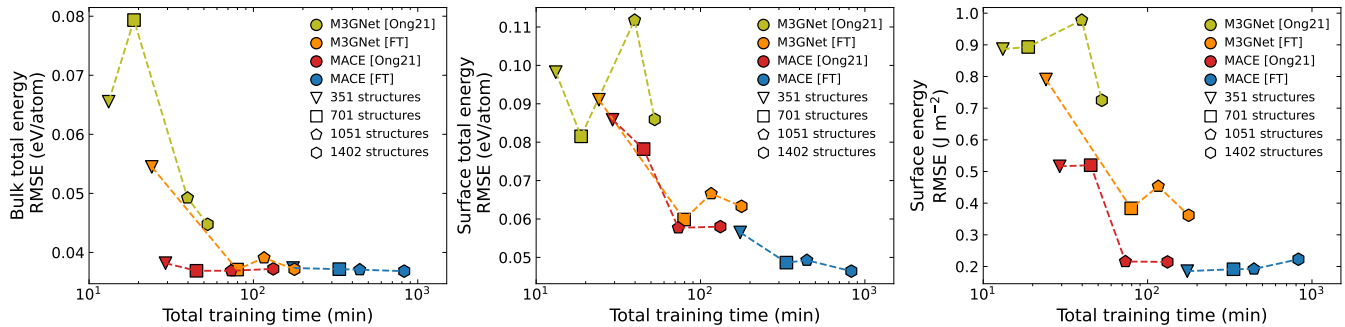

**Figure S6**. Performance and computational cost assessment of fine-tuning and training models from scratch with respect to dataset size. The training was performed using a single NVIDIA Tesla P100 16 GPU. The dataset for fine-tuning and training was the [Ong21] dataset as described in the main text. Downsampling the [Ong21] dataset was performed by randomly selecting structures, and a progressive increase of the data volume incorporates structures from lower volume fine-tunings/trainings.

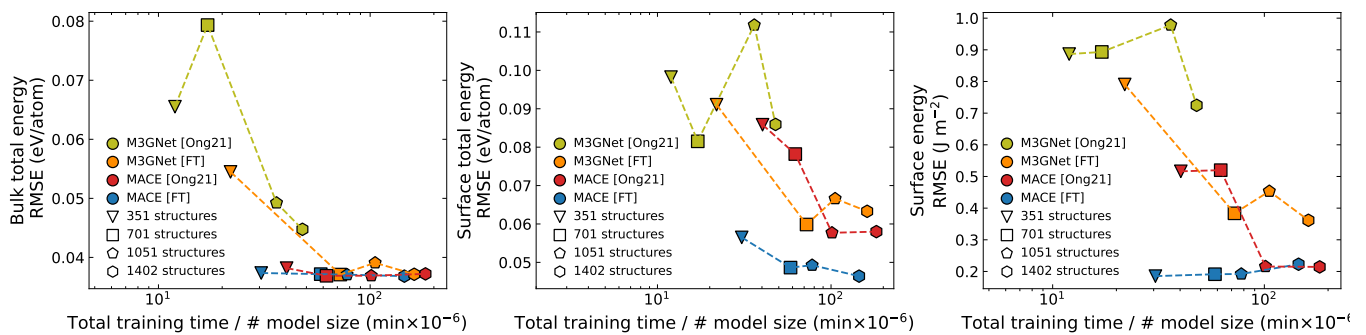

**Figure S7**. Performance and computational cost assessment of fine-tuning and training models from scratch with respect to dataset size considering the size of each model. For MACE [FT] we considered 5.7M parameters which corresponds to the $L = 2$ model, for MACE [Ong21] the number of parameters is 0.724M, for M3GNet [FT] and M3GNet [Ong21] the number of parameters is 1.1M. The training was performed using a single NVIDIA Tesla P100 16 GPU. The dataset for fine-tuning and training was the [Ong21] dataset as described in the main text. Downsampling the [Ong21] dataset was performed by randomly selecting structures, and a progressive increase of the data volume incorporates structures from lower volume fine-tunings/trainings.

## V. HYPERPARAMETER OPTIMIZATION OF SPECIALIZED IPS

### A. MTP

We performed a grid search to select the optimal combination of radius cutoff and the number of polynomial powers. The number of free parameters is determined by the corresponding number of polynomial powers as given by the MLIP package [1]. Our loss function for the grid search is evaluated by the weighted sum of the energy and forces RMSE. The weight for energy contribution is 1.0 while the weight for the forces contribution is 0.01. For fitting the MTP potentials, we fixed the energy weight as 1.0, and the force weight as 0.01. Also, we fixed the size of the radial basis to 8. The maximum number of iterations is set to 500, 1000, and 2000 to ensure the convergence of potentials with 16 or lower polynomial powers, 20 polynomial powers, and 24 polynomial powers, respectively. Our selected hyperparameters are given in Table **S2**.

**Table S2**. Hyperparameters of the MTP model across different chemistries of the MPtrj* dataset.

|                          | Ni   | Cu   | Li   | Mo   | Si   | Ge   |
|--------------------------|------|------|------|------|------|------|
| cutoff radius (Å)        | 4.30 | 5.30 | 4.20 | 4.00 | 5.70 | 4.20 |
| # of polynomial powers   | 16   | 18   | 22   | 24   | 18   | 16   |
| Radial basis size        | 8    | 8    | 8    | 8    | 8    | 8    |
| Energy weight            | 1.00 | 1.00 | 1.00 | 1.00 | 1.00 | 1.00 |
| Force weight             | 0.01 | 0.01 | 0.01 | 0.01 | 0.01 | 0.01 |

## B. NequIP

We performed a grid search to select the optimal combination of radius cutoff, number of features, and maximum $l$ for the representation. We search the radius cutoff between 4.0, 5.0, 6.0, 7.0 Å, the number of features between 32 and 64, and $l_{max}$ between 1 and 2. For constructing NequIP models from [Ong21] and MPtrj* datasets split in a proportion of 90 : 10 for training and validation sets. We used a learning rate of 0.005 and a batch size of 1. The Adam optimizer in the amsgrad mode was used without weight decay. The loss function is based on a weighted sum of energy (`PerAtomMSE` in eV atom$^{-1}$) and force (`ForceMSE` in eV Å$^{-1}$) loss terms set to 4 and 1 respectively.

**Table S3**. Hyperparameters of the NequIP model across different chemistries of the [Ong21] dataset.

|                                  | Ni  | Cu  | Li  | Mo  | Si  | Ge  |
|----------------------------------|-----|-----|-----|-----|-----|-----|
| cutoff radius (Å)                | 5.0 | 5.0 | 7.0 | 5.0 | 6.0 | 7.0 |
| multiplicity of the features     | 64  | 32  | 32  | 32  | 64  | 32  |
| maximum irrep order ($l_{max}$)  | 2   | 2   | 2   | 2   | 2   | 2   |

**Table S4**. Hyperparameters of the NequIP model across different chemistries of the MPtrj* dataset.

|                                  | Ni  | Cu  | Li  | Mo  | Si  | Ge  |
|----------------------------------|-----|-----|-----|-----|-----|-----|
| cutoff radius (Å)                | 6.0 | 7.0 | 6.0 | 7.0 | 6.0 | 6.0 |
| multiplicity of the features     | 64  | 64  | 64  | 32  | 32  | 64  |
| maximum irrep order ($l_{max}$)  | 2   | 2   | 2   | 2   | 2   | 2   |



\end{document}